\documentstyle[12pt]{article}
\def\eettg{\mbox{ $e^+e^-\to t\bar tg$}}
\def\beq{\begin{equation}}
\def\eeq{\end{equation}}
\def\beqa{\begin{eqnarray}}
\def\eeqa{\end{eqnarray}}

\def\tnodd{\mbox{$T_N$-odd}}
\def\tneven{\mbox{$T_N$-even}}

\newcommand{\lsim}{\mathrel{\lower4pt\hbox{$\sim$}}
\hskip-12.5pt\raise1.6pt\hbox{$<$}\;}

\newcommand{\gsim}{\mathrel{\lower4pt\hbox{$\sim$}}
\hskip-12.5pt\raise1.6pt\hbox{$>$}\;}

\begin{document}


\setlength{\textheight}{7.5truein}

\def\vtd{$V_{td}$}
\def\what{\underbar{\hskip.5in}}

\def\lsim{\mathrel{\lower4pt\hbox{$\sim$}}
\hskip-12pt\raise1.6pt\hbox{$<$}\;}

\def\gsim{\mathrel{\lower4pt\hbox{$\sim$}}
\hskip-12pt\raise1.6pt\hbox{$>$}\;}

\begin{titlepage}
\noindent \hspace*{10cm}SLAC-PUB-95-6765\\
\noindent \hspace*{10cm}TECHNION-PH-94-14\\
\rightline{February 1995}
\vfill
\begin{center}
%
%
%
{\bf CP Nonconservation in $e^+e^-\to t\bar tg$}\\
%
%
%
%
%
%
%
%
%
%
%
\vfill
{Shaouly Bar-Shalom$^a$, David Atwood$^b$, Gad Eilam$^a$ and
Amarjit Soni$^c$}\\
\end{center}
\vfill

\begin{flushleft}

a) Department of Physics, Technion, Haifa, Israel.\\
b) Department of Physics, SLAC, Stanford University, Stanford, CA\ \ 94309, USA
\\
c) Department of Physics, Brookhaven National Laboratory, Upton, NY\ \ 11973,
USA\\

\end{flushleft}

\vfill

\begin{quote}
{\bf Abstract}:
CP violation effects in  \eettg\ are examined. CP-odd,
\tnodd\ and \tneven\ observables can both be used to extract information
on the real and imaginary parts of Feynman amplitudes. Two Higgs doublet
mod\-el with CP violating phase from  neutral Higgs exchange is used
to estimate possible effects.
\end{quote}

\vfill

\begin{center}
Submitted to {\it Physics Letters B}
\end{center}

\vfill

\hrule
\vspace{5 pt}
\noindent

* This work was supported in part by the U.S.-Israel Binational Science
Foundation, and by
US Department of Energy
contracts DE-AC03-765F00515 (SLAC)
and DE-AC02-76CH0016 (BNL).
The work of G.E. has been supported in part by the fund for the Promotion
of Research at the Technion.

\end{titlepage}

\eject

%

The prospects for CP violation in top physics has been receiving
considerable attention in the past few years.$^{1\mbox{--}7}$ In the
Standard Model (SM), with its CKM phase, the effects are expected to be
extremely small. However, additional CP violating phases occur rather
naturally in extensions of the SM\null. Besides, non-standard sources of
CP violation may well be a necessity to understand baryogenesis.$^{8}$
Therefore, experimental searches for effects of CP violation are perhaps
the best probes of new physics. Thus it is important to investigate
signatures of the additional phases that may be present. In particular,
due to its mass, the top quark can be a very sensitive probe for the
phase(s) from an extended Higgs sector. Such a phase residing, say, in
neutral Higgs exchanges endows the top quark with a large dipole moment
which drives many interesting CP violation asymmetries in production and
decays of the top quark in high energy reactions.$^{2,3,9}$

In this paper we examine the CP violation effects in the reaction:

\beq
e^-(p_-)+e^+(p_+) \to t(p_t) +\bar t(p_{\bar t}) + g(p_g) \label{emep}
\eeq

\noindent ($g$ is the gluon). The advantage of this simple reaction is
that it allows the possibility of probing both categories of CP
asymmetries that can occur. We recall that CP violating observables can
be subdivided into \tneven\ and \tnodd\ type. Here $T_N$ is the ``naive
time-reversal'' operator (i.e.\ time${}\to-$ time without switching of
initial and final states). \tneven\ observables are driven by imaginary
parts of Feynman amplitudes whereas \tnodd\ observables are proportional
to the real parts of the amplitudes. Of course, being CP violating, both
types of observables do need CP violating phase(s) from the underlying
theory.

CP violation effects can manifest in the momentum distributions
of the incoming and outgoing particles. To illustrate this, let us
define$^{10}$

\begin{center}
\begin{tabular}{lll}
$E  =  \epsilon (p_-,p_+,p_t,p_{\bar t})$; & \quad$s  = 2p_- \cdot p_+$;
& $s_t= (p_t+p_{\bar t})^2$  \\
$u  = (p_--p_+) \cdot (p_t-p_{\bar t})$; & \quad$F  = (p_--p_+) \cdot
(p_t +p_{\bar t})$ &  \\
$G  = (p_-+p_+) \cdot (p_t-p_{\bar t})$ && \\
\end{tabular}
\end{center}

\noindent Any term in the cross section can be expressed by the above
kinematic functions: $G$, $F$ and $E$ are CP-odd, where $G$ and $F$ are
\tneven\ while $E$ is \tnodd. All the other terms are CP-even; thus all
CP violating effects in the momentum distributions will be proportional
to $G$, $F$ or $E$.

Simple examples of \tneven\ observables that can be studied with
reaction (\ref{emep}) are:

\beqa
O_{i1} & \equiv & \vec p_- \cdot (\vec p_t+\vec p_{\bar t})
/s\qquad\qquad (a) \nonumber \\
O_{i2} & \equiv & (E_t-E_{\bar t})/\sqrt{s} \qquad\qquad\; (b) \label{oi123}\\
O_{i3}  &
\equiv & \vec p_g \cdot (\vec p_t-\vec p_{\bar t}) /s \qquad\qquad\; (c)
\nonumber
\eeqa

\noindent We can identify $O_{i1}$ with the forward-backward asymmetry
of the gluon jet while $O_{i2}$ is the energy asymmetry between $t$ and
$\bar t$. The CP violating \tnodd\ observables have to be proportional
to $\epsilon(p_-, p_+, p_t, p_{\bar t})$; since these are the only
independent 4-momenta that are available. This leads us to consider the
following CP-odd, \tnodd\ triple correlation product:

\beqa
O_{r1}  \equiv  \vec p_- \cdot (\vec p_t\times \vec p_{\bar t})/s^{3/2}
\label{or1}
\eeqa

For definiteness we will focus on the effects of a two-Higgs doublet
model (THDM).$^{11,2,12}$ As is well known flavor changing neutral currents
(FCNC) are avoided in such a model by imposing a discrete symmetry. This
then restricts coupling of one Higgs doublet ($\Phi_2$) with charge 2/3
quarks and the other doublet ($\Phi_1$) couples only to the
charge $-1/3$ quarks and the leptons. CP violation is induced in the
model by softly breaking the discrete symmetry in the Higgs potential.
This causes mixing between real and imaginary parts of Higgs fields and
the Higgs mass eigenstates then do not have a definite CP property. The
manifestation of such a CP violation is that the neutral Higgs couple to
fermions with scalar as well as pseudoscalar couplings. Thus

\beq
{\cal L}_{Hff} = H \bar f (a_f + ib_f\gamma_5)f \label{lhff}  \eeq

\noindent where $H$ is the lightest neutral Higgs. For simplicity we are
assuming that the other Higgs particles are heavy enough that their effects can
be
ignored. The $ZZH$ interaction also plays an important role in our
calculation and is  given by

\beq
{\cal L}_{ZZH} = \frac{2m^2_Z}{v} c g_{\mu\nu} Z^\mu Z^\nu \label{lzzh}
\eeq

\noindent where $v=\sqrt{v^2_1 + v^2_2} \equiv (\sqrt{2} G_F)^{1/2}$,
$v_{1,2}$ being the vacuum expectation values of the two Higgs, with the
usual definion of $\tan \beta\equiv v_2/v_1$. In
eqs. \ref{lhff}-\ref{lzzh} the coefficients $a$, $b$, and $c$ are
functions of $\tan \beta$ and the mixing matrix elements between the three
neutral scalar fields.$^{2,12}$

Fig.~1 shows the tree-level Feynman graphs that contribute to the
process $e^+e^-\to t\bar tg$.  Fig.~2 shows the Feynman graphs that (to
one loop order) are calculated to determine the CP asymmetry. From
eqs.~(\ref{lhff}) and (\ref{lzzh}) we see that all the CP violating terms
are proportional to either $ab$ or $cb$.

Before presenting the numerical results we want to augment the list of
CP violating observables. In eqs.~(\ref{oi123}) and (\ref{or1}) we
gave examples of ``naive'' observables. A non-vanishing expectation
value of any one of these would signal CP violation so that experimental
searches for these can be done without recourse to any model. Of course
in a theoretical discussion such as ours one can investigate the
expected asymmetries based on specific models of CP violation. However,
in the context of any given model one can also construct optimal
observables i.e.\ those observables which will be the most sensitive to
CP violation effects in that model. The recipe for such
a construction is very simple. It can be shown$^{12}$ that the optimal
observables of (\tneven\ and \tnodd) CP violation are given by:

\beq
O_{i\rm opt} \equiv \Sigma^{\rm Im}_1/\Sigma_0 \quad , \quad O_{r\rm
opt} \equiv \Sigma^{\rm Re}_1 /\Sigma_0
\eeq

\noindent where the differential cross section (in the variable $\phi$
under consideration) is broken down as:

\beq
\Sigma(\phi) = \Sigma_0(\phi) + \Sigma^{\rm Re}_1(\phi) + \Sigma^{\rm
Im}_1 (\phi)
\eeq

\noindent Here $\Sigma_0(\phi)$ is CP-even piece and $\Sigma_1$ is
CP-odd piece which  is further subdivided into a segment that depends on
the real part of the amplitude and the one that depends on the imaginary
part.

\pagebreak
Let us define $A_O$ to be:

\beq
A_O \equiv <O>/\sqrt{<O^2>}
\eeq

\noindent where $<O^2>$ is the expected variance, then to observe a
non-vanishing
average value $<O>$ with a statistical significance of one sigma one
needs:

\beq
N_{t\bar{t}g} = 1/A^2_O
\eeq

\noindent $N_{t\bar{t}g} = \cal L \sigma$ ($e^+e^-\to t\bar tg$) being the
numbe
of $t\bar{t}g$ events and $\cal L$ is the collider  luminosity.

For definiteness, we will set $a=b=c=1$. Furthermore, we have imposed an
invariant mass cut on the jet pairs so that
$(p_g+p_t)^2$ and $(p_g+p_{\bar{t}})^2 \geq (m_t+m_0)^2$
where we have taken $m_0 = 25$ GeV and $m_t = 174$ GeV.
Also we will first focus
on the case of left-polarized electrons. Fig.~3--4 show our main
numerical results$^{14}$ for $m_H=100$ and 200 GeV respectively. The
number of events needed to see a non-vanishing value (to one sigma) for
some of the naive observables ($O_{i1}$ and $O_{r1}$) and the optimal
observables are shown. We see that near threshold ($E_{CM}\sim400$ GeV)
the \tnodd\ and \tneven\ observables have comparable effectiveness;
\tnodd\ ones are a bit better. As the CM energy increases both types
become worse rather rapidly. However as the CM energy is increased
further the \tneven\ ones improve and can become almost as effective as
they are near threshold. The turn around in energy where they regain
their effectiveness depends on $m_H$. Thus for $m_H=100$ GeV, $E_{CM}
\gsim 500$ GeV is needed and for $m_H=200$ GeV, $E_{CM}\gsim700$ GeV
becomes necessary. Note, though, that the sensitivity of the observables
near threshold does not depend too heavily on the precise value of
$m_H$.

Table~1 gives a brief comparison of the left, right and
un-polarized cases. We note that for the \tneven\ (e.g.\ $O_{i1}$ and
$O_{i{\rm opt}}$) cases the polarization makes a significant difference
and improves their effectiveness by an order of magnitude or even more.
For these it seems that the left-polarized case is marginally better
over the right one. For the \tnodd\ observables (e.g.\ $O_{r1}$ and
$O_{r{\rm opt}}$) beam polarization does not make much of a difference.

We see that $10^5$--$10^6 t\bar tg$ events may be necessary to see an
indication of these effects. We recall that a future $e^+e^-$ collider
could perhaps produce $\sim10^5 t\bar t$ pairs. Table~2 gives the ratio
of $t\bar tg$ to $t\bar t$ events. We note that even at
$\sqrt{s}\gsim800$ GeV the presence of the extra gluon could reduce the
rate by about a factor of 3--4 for the polarized case. Bearing
Figs.~3--4 and Table~2 in mind it would seem that for $e^+e^-\to t\bar
tg$, study at higher energy (i.e.\ away from threshold) would be better,
at least for \tneven\ observables.

It must be emphasized, though, that many simplifying assumptions were
made along the way (e.g.\ $a=b=c=1$) so that the actual size of the
effects could be a lot smaller or even bigger. For example, even a
modest change from the values we used to say $ab\sim3$ (which is
equivalent to setting $\tan \beta = 0.5$)
would increase CP
asymmetries by the same amount (since the asymmetry is linear in $ab$ or
$bc$) and would tend to reduce the number of events from those given in
Fig.~3--4 and Table~1 by about an order of magnitude. Furthermore, the
CP violation may have other sources, say charged Higgs exchanges or
supersymmetry. The key point is that the study of the simple reaction
$e^+e^-\to t\bar tg$, with beam energy of several hundred GeV, via the
observables discussed here, could be a very valuable probe for searching
for non-standard sources of CP violation. It should also be noted that
in this work we have not included the detection of the decay product of
the top quark. In particular, as is well known, the decay of the top
quark acts as an analyzer of the top spin.$^9$ Inclusion of the $t$,
$\bar t$ spins is very likely also to help in the analysis of CP
violation effects. We will return to some of these points in a future
publication.

We would like to thank Dr. G.J. van Oldenborgh for his help in operating
the FF-package for evaluating loop integrals.
This work was supported in part by the U.S.-Israel Binational Science
Foundation, and by DOE contracts DE-AC03-765F00515 and DE-AC02-76CH0016.
The work of G.E. has been supported in part by the fund for the Promotion
of Research at the Technion.
\pagebreak

\bigskip
\noindent{\bf References}
\medskip

\begin{enumerate}

\item C.R. Schmidt and M. Peskin, Phys.\ Rev.\ Lett.\ {\bf69}, 410
(1992); W. Bernreuther and A. Brandenburg, Phys.\ Lett.\ B{\bf314}, 104
(1993); C.R. Schmidt, Phys.\ Lett.\ B{\bf293}, 111 (1992).

\item W. Bernreuthor, T. Schroder and T.N. Pham, Phys.\ Lett.\
B{\bf279}, 389 (1992).

\item A. Soni and R.M. Xu, Phys.\ Rev.\ Lett.\ {\bf69}, 33 (1992).

\item G.L. Kane, G. Ladinski and C.P. Yuan, Phys.\ Rev.\ D{\bf45}, 124
(1992).

\item B. Grzadkowski, Phys.\ Lett.\ B{\bf305}, 384 (1993); B.
Grzadkowski and W-Y. Keung, Phys.\ Lett.\ B{\bf319}, 526 (1993); E.
Christova and M. Fabbrichese, Phys.\ Lett.\ B{\bf320}, 299 (1994).

\item D. Atwood, G. Eilam and A. Soni, Phys.\ Rev.\ Lett.\ 71, 492
(1993).

\item See e.g.\ N. Deshpande {\it et al}., Phys.\ Rev.\ D{\bf45}, 178
(1992); G. Eilam, J. Hewett and A. Soni, Phys.\ Rev.\ Lett.\ {\bf67},
1979 (1991);  B. Grzadkowski and W-Y. Keung, {\it ibid}.

\item For a review see: A.G. Cohen, D.B. Kaplan and A.E. Nelson,
Ann.\ Rev.\ Nucl.\ Part.\ Phys.\ {\bf43}, 27 (1993).
\item D. Atwood, A. Aeppli, and A. Soni, Phys.\ Rev.\ Lett.\ {\bf69},
2754 (1992).

\item D. Atwood, S. Bar-Shalom, and A. Soni, SLAC-PUB-6435 to appear in
Phys.\ Rev.\ D.

\item G.C. Branco and M.N. Rebelo, Phys.\ Lett.\ B{\bf160}, 117 (1985);
J. Liu and L. Wolfenstein, Nucl.\ Phys.\ B{\bf289}, 1 (1987).

\item C.D. Froggett, R.G. Moorhouse, and I.G. Knowles, Nucl.\ Phys.\
B{\bf386}, 63 (1992).

\item D. Atwood and A. Soni, Phys.\ Rev.\ D{\bf45}, 45 (1992).

\item For the FF-package that was used for numerical
evaluation of loop integrals see: G.J. van Oldenborgh,
Comput.\ Phys.\ Commun.\ {\bf66}, 1 (1991).
For the algorithms used in the FF-package
see: G.J. van Oldenborgh and J.A.M. Vermaseren, Z.\ Phys.\ C{\bf46}, 425
(1990).
\end{enumerate}
\pagebreak

\bigskip
\noindent {\bf Figure Captions}
\medskip

\begin{description}

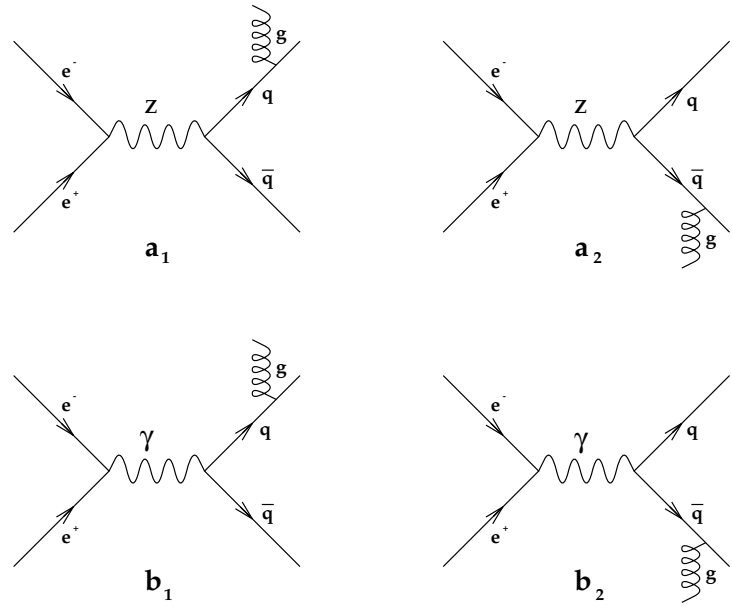
\item[\rm Fig.\ 1] Tree-level Feynman diagrams contributing to
$e^+e^-\to t\bar tg$.

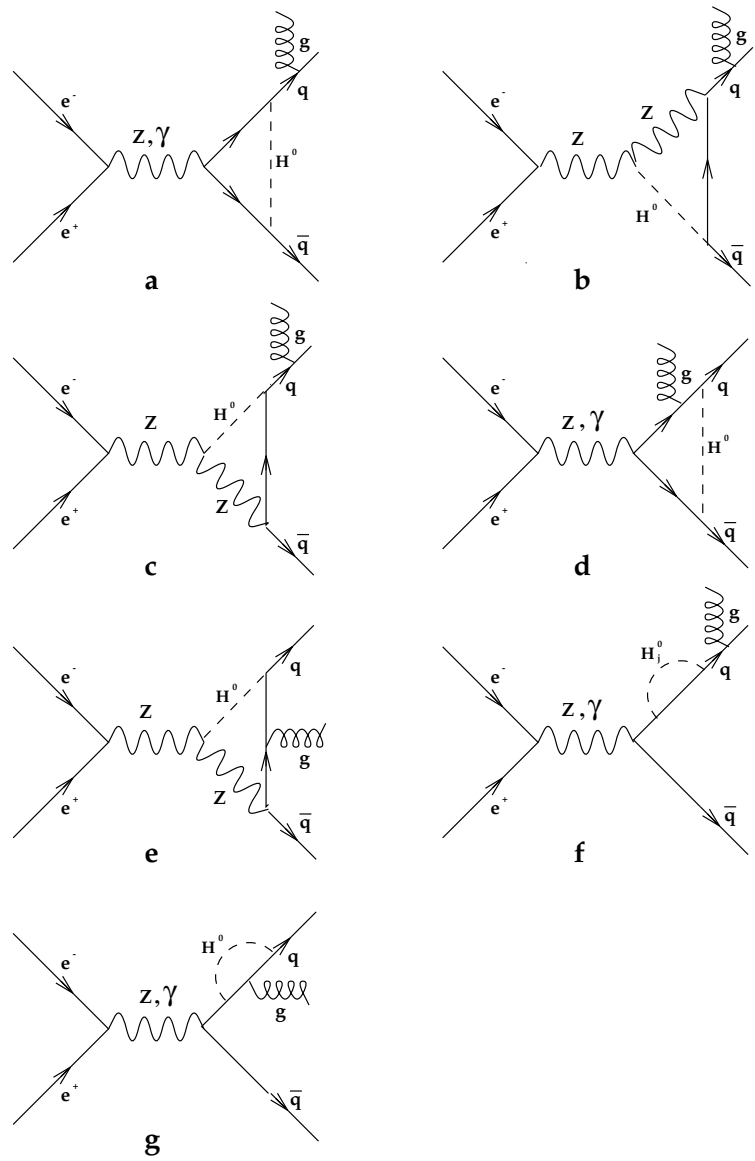
\item[\rm Fig.\ 2] CP violating Feynman diagrams contributing to \eettg\
to one loop order in a two Higgs doublet model. Diagrams with permuted
vertices (e.g. $t \rightarrow \bar{t}$) are not drawn.

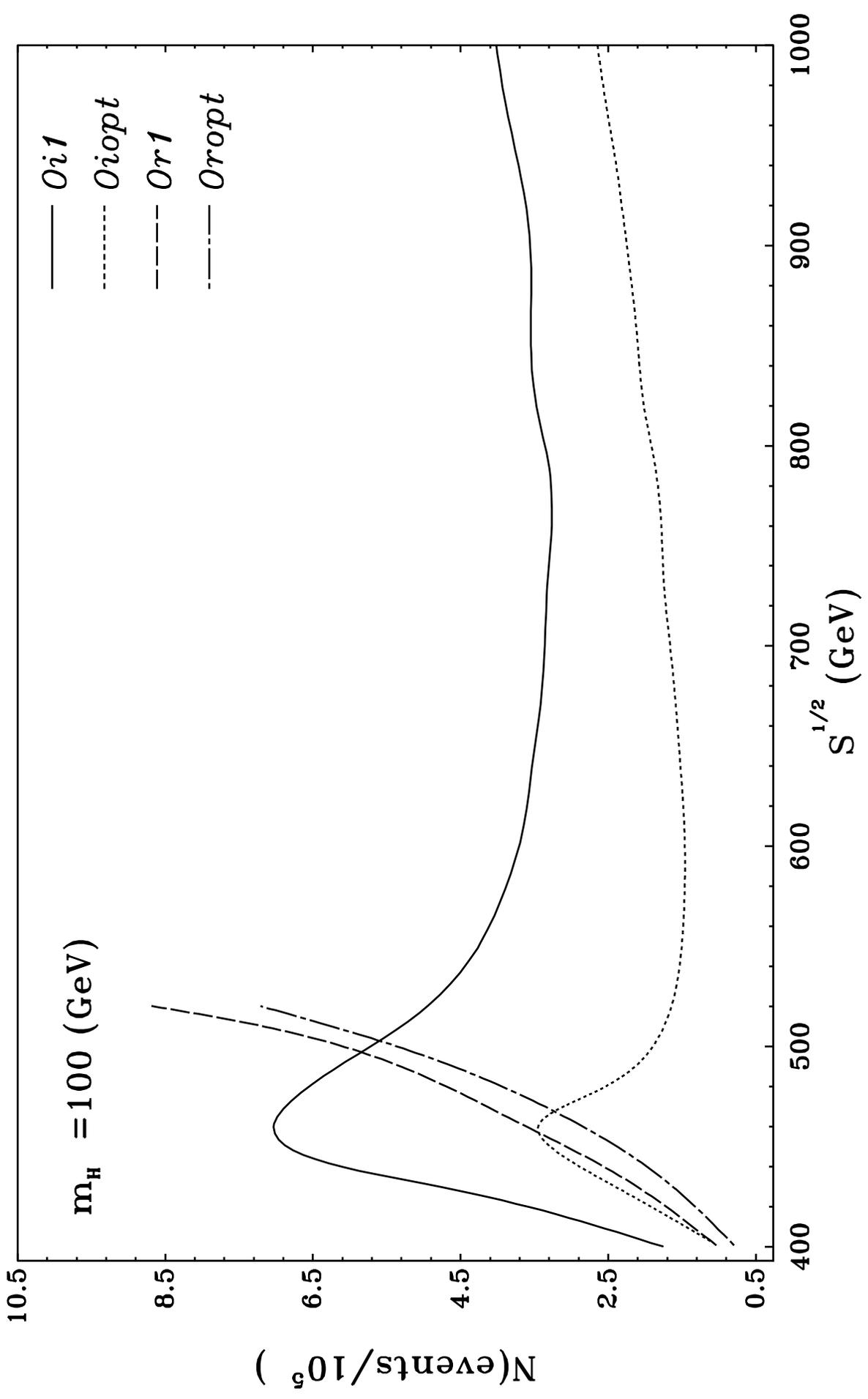
\item[\rm Fig.\ 3] Number of events (in units of $10^5$) needed to
detect CP violation via $\langle O_{i1}\rangle$, $\langle
O_{r1}\rangle$, $\langle O_{i{\rm opt}}\rangle$ and $\langle O_{r{\rm
opt}}\rangle$ as a function of total beam energy, $m_H=100$ GeV is used.

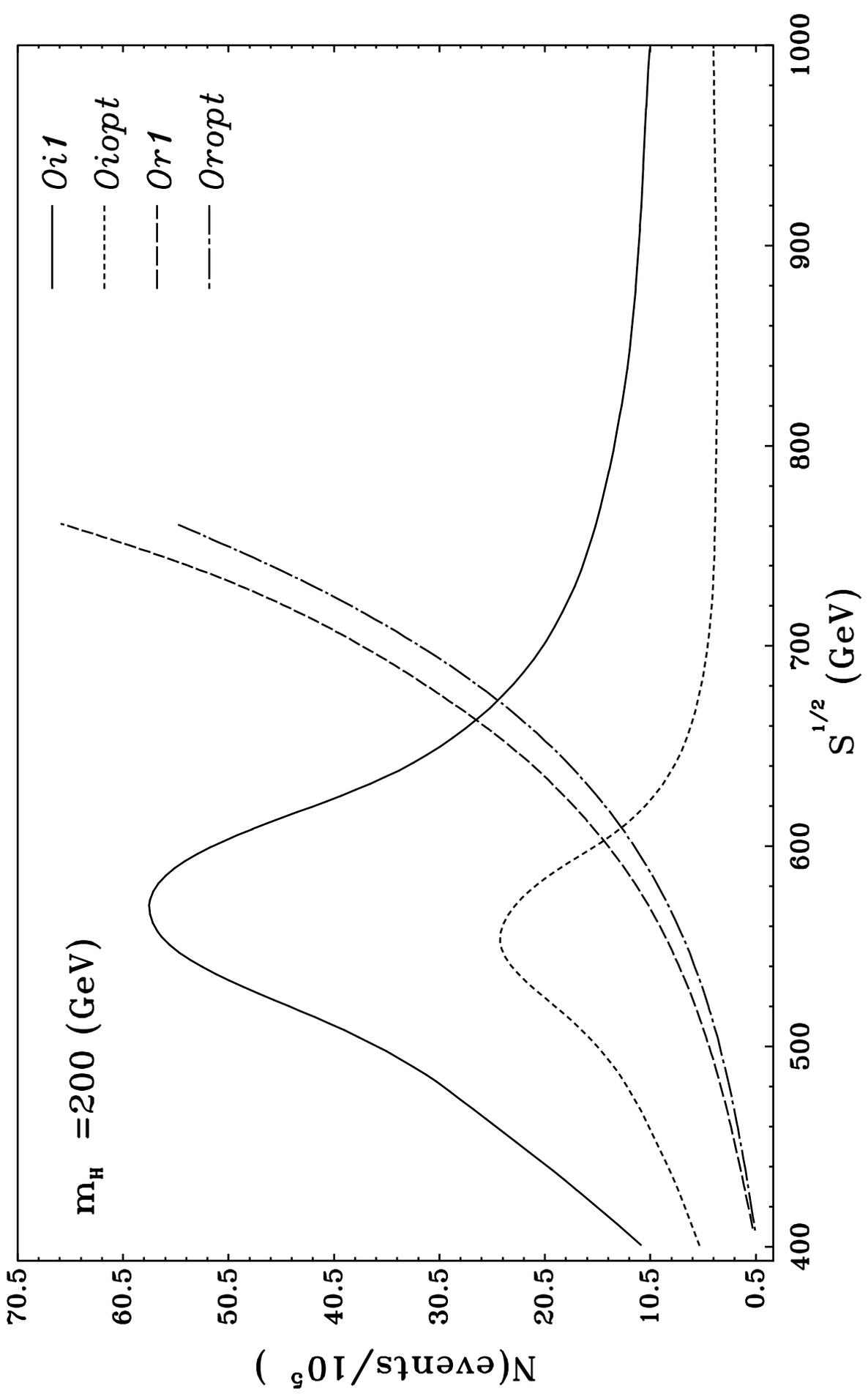
\item[\rm Fig.\ 4] Same as Fig.\ 2 except $m_H=200$ GeV\null.

\end{description}

\begin{table}[htbp]
\caption[entry]{ The unpolarized case (pol${}=0$) is compared with left
polarization (pol${}=-1$) and the right polarization (pol${}=+1$).The number of
events in units of $10^5$ needed, for detection of asymmetries, to one sigma
are given.
 The values of $\sqrt{s}$ and $m_H$ are given in GeV.}
\bigskip
\moveleft1in\vbox{
\begin{tabular}{|l|r|c|c|c|c|c|c|c|c|}
\hline
& & \multicolumn{2}{c|}{$O_{i1}$} & \multicolumn{2}{c|}{$O_{r1}$} &
\multicolumn{2}{c|}{$O_{i{\rm opt}}$} & \multicolumn{2}{c|}{$O_{r{\rm
opt}}$} \\
\cline{3-10}
\protect\footnotesize$\sqrt{s}$ & \protect\footnotesize pol. &
\protect\footnotesize$m_H=100$ &
\protect\footnotesize$m_H=200$ & \protect\footnotesize$m_H=100$ &
\protect\footnotesize$m_H=200$ &  \protect\footnotesize$m_H=100$ &
\protect\footnotesize$m_H=200$ & \protect\footnotesize$m_H=100$ &
\protect\footnotesize$m_H=200$ \\
\hline  & -1 & 1.8  & 11.5  & 1.0  & 0.9  & 1.0  & 6.0  & 0.8  & 0.7 \\
\cline{2-10}
400 & 0 & 22.5  & 134.8  & 1.8  & 1.5  & 6.5  & 37.0  & 0.7  & 0.6  \\
\cline{2-10}
& 1 & 2.3  & 17.1  & 0.7  & 0.6 & 1.3 & 8.4 & 0.6  & 0.5  \\
\hline
\multicolumn{10}{|c|}{} \\
\hline  & -1 & 3.4  & 20.0  & 66.7  & 37.8  & 1.7  & 5.1  & 57.2 & 32.9  \\
\cline{2-10}
700 & 0 & 48.6  & 263.9  & 124.6  & 70.2  & 12.2  & 38.6  & 52.8  & 30.3  \\
\cline{2-10}
& 1 & 4.5 & 30.8  & 55.1  & 30.7  & 2.0  & 5.9 & 45.0  & 25.9  \\
\hline
\multicolumn{10}{|c|}{} \\
\hline  & -1 & 4.0  & 10.5  & 625.6  & 363.8  & 2.6  & 4.5  & 496.9  & 320.4
\\
\cline{2-10}
1000 & 0 & 63.4  & 158.2  & 1193.9 & 699.8  & 20.5  & 35.3  & 461.9  & 301.2
\\
\cline{2-10} & 1 & 5.0  & 14.0  & 547.9  & 325.7  & 3.1  & 5.4  & 399.2
& 264.6  \\
\hline
\end{tabular}
}
\bigskip\bigskip
\bigskip\bigskip

\noindent Table 2: The numbers for the ratio: $\left[\frac{\sigma(e^+e^-\to
t\bar tg)}{\sigma(e^+e^-\to t\bar t)} \right]$ with the cut: $E_{\rm glue} \ge
25$ [GeV] and for different polarizations. The value of $\sqrt{s}$ is given in
GeV.
\bigskip

\begin{tabular}{|l|c|c|c|c|c|}
\hline
$\sqrt{s}\Rightarrow$ & & & & & \\
\cline{1-1}
$\Downarrow$ Pol. & 400 & 550 & 700 & 850 & 1000 \\
\hline
Left (-1) & $0.003$ & 0.08 & 0.19 & 0.29 & 0.49 \\
\hline
Unpol. (0) & $0.007$ & 0.19 & 0.42 & 0.63 & 0.83 \\
\hline
Right (1) & $0.006$ & 0.13 & 0.27 & 0.40 & 0.51 \\
\hline
\end{tabular}
\end{table}
\pagebreak

\end{document}